# Fermi Surface of Bi2212: a Systematic Revisit and Identification of Almost Perfectly Nested Fermi Surface Segments

D.L. Feng[1], W.J. Zheng[1], K.M. Shen[1], D.H. Lu[1], F. Ronning[1], J.-i. Shimoyama[2], K. Kishio[2], G. Gu[3], Dirk Van der Marel[4], Z.-X.Shen[1]

[1] Department of Physics, Applied Physics and Stanford Synchrotron Radiation Laboratory, Stanford University, Stanford, CA 94305, USA
[2] Department of Applied Chemistry, University of Tokyo, Tokyo, 113-8656, Japan
[3] School of Physics, University of New South Wales, P. O. Box 1, Kensington, NSW, Australia 2033
[4] Material Science Center, University of Groningen, Nijenborgh 4, 9747 AG Groningen, The Netherlands

(October 14, 1999)

The Fermi surface of the Bi2212 system is systematically studied using a variety of photon energies. In addition to reconciling the conflicting reports on the Fermi surface of this important material, we identify almost perfectly nested Fermi surface segments parallel to the (1,0) or (0,1) direction. This general yet long overlooked feature of the Fermi surface is intriguingly similar to the Fermi surface of the $La_{1.48}Nd_{0.4}Sr_{0.12}CuO_4$ system.

After almost a decade of angle resolved photoemission spectroscopy (ARPES) studies on the $Bi_2Sr_2CaCu_2O_{8+y}$ (Bi2212) system, the Fermi surface (FS) of this system remains controversial [1–6]. Earlier data suggest the presence of two FS's, with one centered at $(0,0)$ and the other at $(\pi,\pi)$ [2]. On the other hand, researchers [4] have constructed the following physical picture for the FS of the Bi2212 system: a large, round, hole-like pocket centered at $(\pi,\pi)$, with its enclosed volume corresponding to the density of doped holes [3]. Furthermore, this FS can be fit very well by a simple tight binding model of a single plane, despite the fact that the high temperature superconductors are known to be strongly correlated and the bilayer effect is expected from calculation and is observed in other compounds. This latter picture, which has been gaining acceptance, is nevertheless challenged by even more recent data from higher photon energy that only show one piece of FS centered at $(0,0)$ [5,6].

In most of these studies, the FS is determined by either fitting the ARPES peak position over several emission angles and extrapolating the dispersion to the Fermi energy $E_F$ (dispersion method) [1–3,5], or looking for the maximum ARPES intensity within a small window around $E_F$ (max-$A(\vec{k}, E_F)$ method) [4,6]. However, these methods suffer some significant shortcomings. In particular, the dispersion method requires certain subjectivity, and the max-$A(\vec{k}, E_F)$ method has some systematic error [7]. Furthermore, both of these methods suffer from complications related to the superstructure, a problem specific to the Bi2212 system.

A third method, namely the $\nabla n$ method, which is based on the momentum distribution of the density of states $n(\vec{k})$, has been successfully applied to map the FS's of $Cu$, $TiTe_2$ [8], $YBa_2Cu_3O_{7-x}$ [9], $Sr_3Ru_2O_7$ [10], and $Ca_2CuO_2Cl_2$ [11]. In this method, one finds the maximum of $|\nabla_{\vec{k}} n(\vec{k})|$, since the Fermi vector is defined as the location of the discontinuity or steepest descent of $n(\vec{k})$. Under the sudden approximation [12,13], $n(\vec{k})$ can be obtained using ARPES data $n(\vec{k}) = \int A(\vec{k}, \omega) f(\omega) d\omega$, where $A(\vec{k}, \omega)$ is the spectral function and $f(\omega)$ is the Fermi function. Because of the large frequency integration window, the $\nabla n$ method is less susceptible to the superstructure effect, which is particularly important in the Bi2212 system. Moreover, the $\nabla n$ method is especially suitable for a systematic unbiased study.

This paper reports experimental results using the $\nabla n$ method, which is further complemented by the max-$A(\vec{k}, E_F)$ method and the traditional dispersion method. We found two types of Fermi surface features that are emphasized at various photon energies. At 32.3 $eV$ photon energy, our data show a Fermi surface feature centered at $(0,0)$, consistent with the recent data recorded at similar photon energy [5]. At 22.4 and 55 $eV$ photon energy, on the other hand, we found a Fermi surface feature that is similar to a pocket at $(\pi,\pi)$. The presence of these two kinds of Fermi surface features in the *same sample* reconciles conflicting reports in the literature [2–6]. More importantly, this systematic approach allows us to identify strikingly straight FS segments that are parallel to the $(1,0)$ or $(0,1)$ direction, and these segments are off of the $(1,0)$ or $(0,1)$ line by about $\pi/4$. This feature is intriguingly similar to that of the charge ordered $La_{1.48}Nd_{0.4}Sr_{0.12}CuO_4$ system.

The measurements were conducted at Beamline V-3 of the Stanford Synchrotron Radiation Laboratory. The optimally doped sample and the overdoped sample have $T_C$ of 90 $K$ and 79 $K$ respectively. The angular resolution was ±1 degree and the overall system energy resolution, including both the beamline and the analyzer, was 35 $meV$, 47 $meV$, and 80 $meV$ for the photon energies of 22.4 $eV$, 32.3 $eV$, and 55 $eV$ respectively in the measurements. The integration windows for measuring the $n(\vec{k})$ and $A(\vec{k}, E_F)$ were set to $[-500\ meV, 100\ meV]$ and $[-100\ meV, 100\ meV]$ relative to $E_F$ respec-tively. The samples were cleaved *in situ*, and the chamber pressure was better than $5 \times 10^{-11}$ *torr* for all the measurements. Data were taken in the $\vec{k}$-space octant $\Gamma(0,0) - \bar{M}(\pi,0) - Y(\pi,\pi) - \Gamma(0,0)$ at a temperature of 100 $K$. Valence band spectra taken at the beginning and the end of the experiments showed little difference, indicating negligible sample aging.

In reality, the photoemission cross section is not a constant but a function of electron and photon momentum and energy, causing a modulation of the $n(\vec{k})$ data obtained from ARPES spectra. At a given photon energy, the cross section $M(\vec{k}, \omega)$ can be roughly approximated as $M(\vec{k}, \omega) \propto |\vec{A} \cdot \vec{k}|^2 \propto k^2 \cos^2 \theta$, where $\vec{A}$ is the vector potential of the photon field and $\theta$ is the angle between $\vec{k}$ and $\vec{A}$, along the (1, 0) direction in our case. To illustrate this so-called matrix element effect, Fig.1a shows the numerical $n(\vec{k})$ calculated from a tight binding model for the Bi2212 system at finite temperature, while Fig.1b shows the same $n(\vec{k})$ multiplied by $M(\vec{k}, \omega)$. Clearly, Fig.1b, or the $n(\vec{k})$ that is experimentally measured, is very different from the true momentum distribution; it is severely suppressed near $(0, 0)$ point and the $(0, 0) - (\pi, \pi)$ cut [14]. However, the FS itself is not altered by the matrix element effects.

The two-fold symmetrized color scale plot of $n(\vec{k})$ taken from the overdoped sample using 22.4 $eV$ photon energy is shown in Fig.1c, from which one can clearly observe the matrix element induced suppression as described in Fig.1b. The color scale plot of $|\nabla_{\vec{k}} n(\vec{k})|$ is shown in Fig.1d. By definition of the $\nabla n$ method, the locus of local maxima in the $|\nabla_{\vec{k}} n(\vec{k})|$ plot is the FS, denoted by the hashed lines in $n(\vec{k})$ plots; the extra structure is due to the matrix element effects. The Fermi vectors determined by the conventional dispersion method are shown by solid circles overlaid on the $n(\vec{k})$ plot. These two Fermi surfaces overlap well within the experimental angular resolution, further confirming the applicability of the $\nabla n$ method in the Bi2212 system. The most striking feature of this FS is that it possesses two long straight sections at $(\pi, (0.25 \pm 0.05)\pi)$ to $(0.5\pi, (0.25 \pm 0.05)\pi)$ and by symmetry $((0.25 \pm 0.05)\pi, \pi)$ to $((0.25 \pm 0.05)\pi, 0.5\pi)$. This strikingly straight FS feature has not been explicitly identified before, although it is presented in the earlier data [2]. The reasons that this feature has long been overlooked are likely because no global $\nabla n$ method had been previously attempted on the Bi2212 system, data points were not sampled densely enough in the momentum space, and that sample quality was poor in some earlier studies.

Fig.2 shows experimental data using photons of 22.4 $eV$, 32.3 $eV$, and 55 $eV$ from the optimally doped sample. The 22.4 $eV$ $n(\vec{k})$, $|\nabla_{\vec{k}} n(\vec{k})|$, and $A(\vec{k}, E_F)$ data are shown in panel (a1), (b1), and (c1) respectively. One again sees the strikingly straight Fermi surface segment in the $|\nabla_{\vec{k}} n(\vec{k})|$ data. The $\nabla n$ result is complemented by the result of the max-$A(\vec{k}, E_F)$ method. Although the max-$A(\vec{k}, E_F)$ result in (c1) has signal other than the straight segment identified in (b1) because of the flat band dispersion in this region and the limited energy resolution, the dispersion method confirmed the FS shape in (b1), just like Fig.1c. We note that although the locus of local maxima of the $|\nabla_{\vec{k}} n(\vec{k})|$ in (b1) is very straight, the $n(\vec{k})$ plot in (a1) shows some curvature. As the $n(\vec{k})$ contour plot is that of constant $n(\vec{k})$, it can differ slightly from the maximum $|\nabla_{\vec{k}} n(\vec{k})|$. From the 32.3 $eV$ data in (a2), (b2), and (c2), one sees a very different picture. Both the $\nabla n$ method and the max-$A(\vec{k}, E_F)$ method give a Fermi surface that resembles a pocket centered at $(0, 0)$ without the straight FS segment and the flat band near $(\pi, 0)$. Our max-$A(\vec{k}, E_F)$ method data confirm those reported recently using the same photon energy [5,6], although no map of $n(\vec{k})$ and its derivative have been reported before. It should be noted that the max-$A(\vec{k}, E_F)$ result is slightly shifted from the $\nabla n$ result; this small effect is caused by a small systematic error in the max-$A(\vec{k}, E_F)$ method and its higher requirement of the energy resolution [7]. From 55 $eV$ data in (a3) and (b3), one again sees the straight FS segment similar to those seen at 22.4 $eV$. The extra feature in (b3) is probably an artifact due to termination of the data sampling points and enhanced superstructure at this photon energy.

The two Fermi surface pieces seen at different photon energies are quite consistently observed. A natural question is whether there is only one piece of Fermi surface centered at $(\pi, \pi)$ [3], and the other feature seen at 32.3 $eV$ is due to superstructure effects [15]. This is unlikely, because one can see the superstructure effects in Fig.2(c1) and (c2). The key here is that the main Fermi surface intensity is so strong that it is easily discernible from the weaker superstructure signal. If there is only one Fermi surface and the other piece is due to the superstructure, this effect should not depend on the photon energy. The second question is whether the piece of Fermi surface seen at 32.3 $eV$ is a result arising from matrix element or final state effects. This is not the case either, as the Fermi surface features are checked by three different methods: the traditional dispersion method, the max-$A(\vec{k}, E_F)$ method, and the $\nabla n$ method, which are complementary to each other. The observation of the same straight Fermi surface segment at 55 $eV$ shows that it is not a peculiar effect at low photon energy. The third possibility is that the two FS features seen at different photon energies are caused by $k_z$ dispersion of the same feature. Since the bands are rather flat near $(\pi, 0)$, a small dispersion can cause a significant change to the shape of the FS. In this scenario, the two Fermi surface pieces seen are two cross-sections at specific $k_z$, and the Luttinger volume is compensated by variation with $k_z$. On the other hand, this scenario also implies that one could see FS features with its shape in between those observed at 22.4 $eV$ and 32.3 $eV$ at some other photon energies. However, no other FS feature has been observed, instead, at 55 $eV$, one sees the same FS feature as that observed at 22.4 $eV$. This is an issue that needs more investigation. Taking together the data here and those in the literature, we suggest the presence of two kinds of FS features, as in Fig.3, where the two experimental Fermi surface features are plotted together. One is the piece centered at $(0,0)$ while the other is the piece with strikingly straight FS segments. It is presently unclear whether the two pieces we see are due to $k_z$ dispersion or due to different photon energies emphasizing different Fermi surface pieces. However, the observation of the previously controversial results in the same sample clarifies the current situation.

The presence of the strikingly straight Fermi surface segments within our experimental resolution deserves some more attention, as it is possibly the most concrete new finding of our experiment and the data from the two samples in Fig.1 and Fig.2 both show it very clearly. This is a robust feature in the Bi2212 system as we have seen it in a wide range of doping [16]. Such a long straight Fermi surface segment is not expected from band theory. However, this feature is intriguingly similar to that seen in 1/8 doped $La_{1.48}Nd_{0.4}Sr_{0.12}CuO_4$ in which static spin and charge ordering have been reported [17–19]. In the $La_{1.48}Nd_{0.4}Sr_{0.12}CuO_4$ case, the straight "Fermi surface" feature at $\pi/4$ is probably associated with 1/4 filled stripes in a local description of the electronic structure [20]. On the other hand, the straight FS shape would also yield some intriguing numbers in the context of nesting and charge density wave (CDW), although it is not consistent with the stripe interpretation and is less applicable to $La_{1.48}Nd_{0.4}Sr_{0.12}CuO_4$. Taking the width between two straight Fermi surface segments as a fraction $\nu$ of the entire Brillouin zone (BZ), the occupied part of the BZ is then approximately $N = 2\nu - \nu^2$. The number of electrons per unit cell equals $2N$ due to spin degeneracy, and the doping of the cuprates is counted from half filling ($N = 1$), hence $x = 1 - 2N = 1 - 4\nu + 2\nu^2$. When the Fermi surface nesting is commensurate with the lattice for the CDW instability, $\nu = \frac{1}{n}$ with $n$ being an integer. We have $x = (2 + n^2 - 4n)/n^2$ and the following series of 'magic' fractions is obtained: $x = \frac{1}{8}, \frac{7}{25}, \frac{7}{18}, \frac{23}{49}...$ for hole doping. So far only crystals with the magic doping fractions $\frac{1}{8}$ and $\frac{7}{25}$ have been prepared in the hole doped materials. Interestingly, superconductivity is suppressed at $\frac{1}{8}$ doping and terminates near $\frac{7}{25}$ (0.28) doping.

In the case of Bi2212, no static CDW has been reported and quasiparticle like peaks are observed, and one is torn between two very different pictures. On the one hand, the quasiparticle and the Fermi surface nesting picture seems reasonable. The doping levels of the samples we studied are higher than 1/8, however, a feedback of soft fluctuations on the electronic bands is expected if the Fermi surface is close to perfect nesting. Static order may set in when the nesting is locked to the lattice. It is currently not clear how the piece of Fermi surface centered at $(0,0)$ would fit into this context. Since this piece of Fermi surface is wider than the $\vec{k}$ resolution near $(\pi, 0)$, unlike the other piece, we cannot be sure whether it is as strongly nested; the data in Fig.2(c2) also indicate some curvature. Within the framework of the band-like description in $\vec{k}$-space, we cannot understand it in ways other than the bonding-antibonding splitting picture discussed earlier [2], or some scenario involving $k_z$ dispersion. On the other hand, one has to ask that if the straight Fermi surface feature looks so similar to those seen in the stripe phase, how can they stem from very different physics. It is possible that if one starts with a two-component model that consists of 1D 1/4 filled charge stripes and nearly 1/2 filled 2D domains, one may be able to model the two components. This idea is consistent with our observation that the maximum $n(\vec{k})$ pattern shrinks away from $(\pi, 0)$ when underdoping makes the data more similar to that of the undoped insulator, plus the observed straight Fermi surface segments [16]. In this context, one sees a complex picture regarding the electronic structure of Bi2212. One has the straight Fermi surface segment and the feature near $(\pi, 0)$ region that resembles those in Nd-LSCO, where the stripe interpretation is the most reasonable. However, one also has the well-defined peak near the $E_F$ along $(\pi, \pi)$ direction, a result not easily reconcilable with stripes. We leave this as a challenge for theory.

To summarize, we have re-examined the Fermi surface issue of the Bi2212 system, reconciling the conflicting results reported recently. Our systematic study has identified long straight Fermi surface segments that resembles those seen in the charge ordered $La_{1.48}Nd_{0.4}Sr_{0.12}CuO_4$.

The experimental data were recorded at the Stanford Synchrotron Radiation laboratory which is operated by the DOE Office of Basic Energy Science Division of Chemical Sciences and Material Sciences. The Material Sciences Division also provided support for the work. The Stanford experiments are also supported by the NSF grant 9705210 and ONR grant N00014-98-1-0195-A00002.

[1] C. G. Olson et al., Science **245**, 731 (1989); Phys. Rev. B **42**, 381 (1990).
[2] D. Dessau et al., Phys. Rev. Lett. **71**, 2781 (1993); Z.-X. Shen and D. Dessau, Phys. Rep. **253**, 1 (1995)
[3] H. Ding et al., Phys. Rev. Lett. **76**, 1533 (1996); Phys. Rev. Lett. **78**, 2628 (1997).
[4] P. Aebi et al., Phys. Rev. Lett. **72**, 2757 (1994).
[5] Y.-D. Chuang et al., cond-mat/9904050.
[6] N. L. Saini et al., Phys. Rev. Lett. **79**, 3467 (1997).
[7] L. Kipp et al., cond-mat/9905400.
[8] Th. Straub et al., Phys. Rev. B **55**, 13473 (1997).
[9] M. C. Schabel et al., Phys. Rev. B **57**, 6107 (1998). A

more detailed comparison between different FS extraction methods are also discussed there.

[10] A. V. Puchkov et al., Phys. Rev. B **58**, 6671 (1998).
[11] F. Ronning et al. Science **282**, 2067 (1998).
[12] S. Hufner, *Photoemission Spectroscopy*, Springer-Verlag, New York, (1995)
[13] M. Randeria et al., Phys. Rev. Lett **74**, 4951 (1995).
[14] The supression of $n(\vec{k})$ near $(0,0)$ also comes from the limited energy window we can apply as the many-body interaction pushes the spectral weight to very high energy, but we will focus on the polarization effect here for simplicity.
[15] J. Mesot et al., Phys. Rev. Lett **82**, 2618 (1999).
[16] D. L. Feng et al., in preparation.
[17] J. M. Tranquada et al., Nature **375**, 561 (1995).
[18] M. V. Zimmermann et al., Europhys. Lett **41**, 629 (1998).
[19] X.J Zhou et al., Science **286**, 268 (1999).
[20] V. J. Emery, et al. Phys. Rev. Lett. **64**, 475 (1990); Phys. Rev. B **56**, 6120 (1997); J. Zaanen, et al. Phys. Rev. B **40**, 7391 (1989).

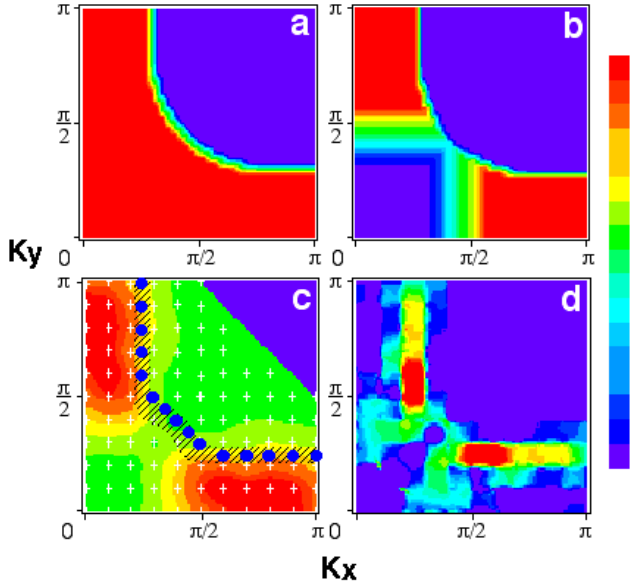

FIG. 1. (color) Scale plots of (a) $n(\vec{k})$ from tight-binding band calculations; (b) same as (a) but with photoemission cross section considerations; (c) $n(\vec{k})$ (with white crosses representing the sampled $\vec{k}$ points) of the overdoped sample ($T_C \sim 79\ K$). The Fermi vectors determined by the dispersion method are shown as the blue solid circles, and the FS determined by the $\nabla n$ method is shown as the hashed line region; and (d) $|\nabla_{\vec{k}} n_{\vec{k}}|$ for data shown in (c). The color bar on the right indicates the linear color scales of the shown quantities, which are always in arbitrary units throughout the paper. The lattice constant $a$ is set to 1 for simplicity and convention; as a result, $\vec{k}$ is unitless.

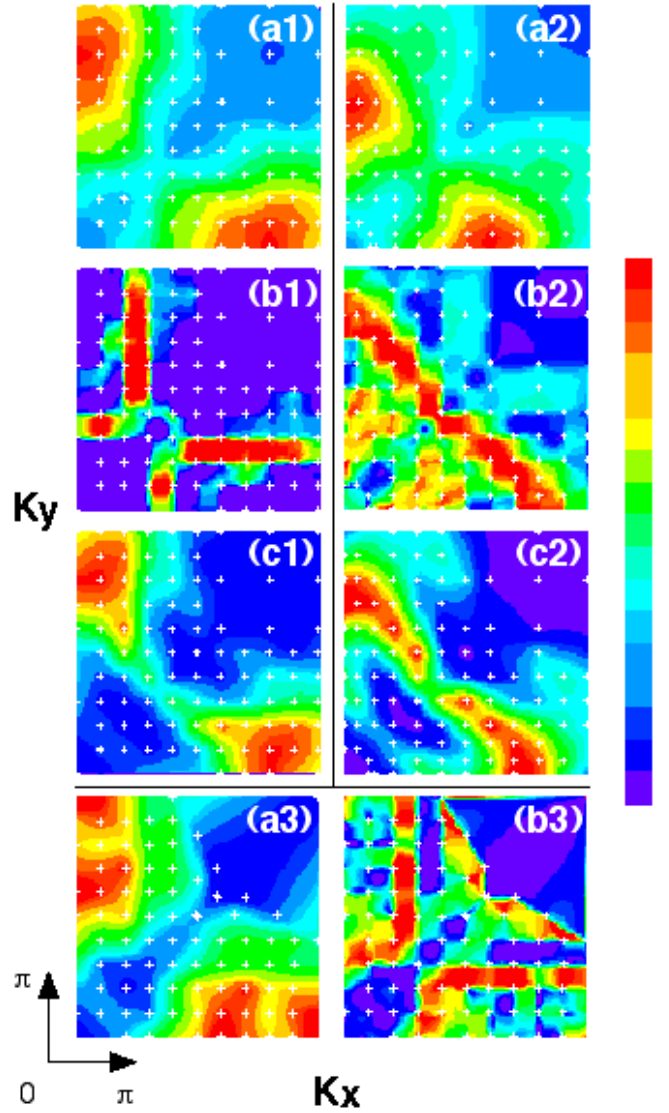

FIG. 2. (color) Two-fold symmetrized scale plots of the optimally doped sample ($T_C \sim 90\ K$) with the white crosses representing the sampled $\vec{k}$ points. (a1), (b1), and (c1) show $n(\vec{k})$, $|\nabla_{\vec{k}} n_{\vec{k}}|$, and $A(\vec{k}, E_F)$ respectively for data taken at 22.4 $eV$ photon energy. (a2), (b2), and (c2) show $n(\vec{k})$, $|\nabla_{\vec{k}} n_{\vec{k}}|$, and $A(\vec{k}, E_F)$ respectively for data taken at 32.3 $eV$ photon energy. (a3) and (b3) show $n(\vec{k})$ and $|\nabla_{\vec{k}} n_{\vec{k}}|$ respectively for data taken at 55 $eV$ photon energy. The ranges of $k_x$ and $k_y$ are both from 0 to $\pi$ for all eight panels.

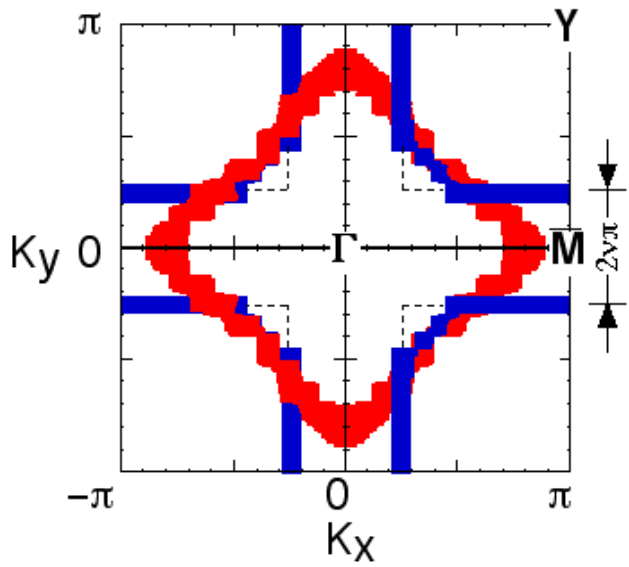

FIG. 3. (color) Eight-fold symmetrized experimental Fermi surface as derived from the local $|\nabla_{\vec{k}} n_{\vec{k}}|$ maxima locus in Fig.2(b1) (blue) and Fig.2(b2) (red). The distance between the nesting pieces is denoted as $2\nu\pi$. The dashed lines are the hypothetical Fermi surface discussed in the text.